\documentclass[preprint]{aastex}
\usepackage{epsfig}

\shorttitle{Burud et al.}
\shortauthors{Optical time delay measurement of  B1600+434}

\begin{document}

   \title{An optical time-delay estimate for the double gravitational lens system B1600+434
             \thanks{Based on observations made with the Nordic Optical 
	      Telescope,
              operated on the island of La Palma jointly by Denmark, Finland,
              Iceland, Norway, and Sweden, in the Spanish Observatorio del
              Roque de los Muchachos of the Instituto de Astrofisica de
              Canarias.}}

\author{I. Burud}
\affil{Institut d'Astrophysique et de G{\'e}ophysique de Li{\`e}ge,
Avenue de Cointe 5, B-4000 Li{\`e}ge, Belgium}
\author{J. Hjorth}
\affil{Astronomical Observatory, University of Copenhagen, 
Juliane Maries Vej 30, DK--2100~Copenhagen {\O}, Denmark}
\author{A. O. Jaunsen}
\affil{Institute of Theoretical Astrophysics, University of Oslo,
Pb.~1029 Blindern, N0315~Oslo, Norway}
\author{M. I. Andersen, H. Korhonen}
\affil{Division of Astronomy, University of Oulu, P.O. Box 3000, FIN--90014 Oulun Yliopisto, Finland}
\author{J. W. Clasen}
\affil{Nordic Optical Telescope, Apartado 474, E--38700 St.~Cruz de La Palma,
Canary Islands, Spain}
\author{J. Pelt}
\affil{Tartu Astrophysical Observatory, T\~{o}ravere, 61602, Estonia}
\author{F. P. Pijpers}
\affil{Theoretical Astrophysics Center, University of Aarhus, DK--8000~\AA rhus C, Denmark}
\author{P. Magain}
\affil{Institut d'Astrophysique et de G{\'e}ophysique de Li{\`e}ge,
Avenue de Cointe 5, B-4000~Li{\`e}ge, Belgium}
\and
\author{R. {\O}stensen}
\affil{Department of Physics, University of Troms{\o}, Troms{\o}, Norway}



\begin{abstract}
We present optical $I$-band light curves of the gravitationally lensed
double QSO B1600+434 from  observations obtained at the Nordic Optical
Telescope (NOT)  between April 1998 and November  1999. The photometry
has  been performed  by  simultaneous deconvolution  of  all the  data
frames, involving  a numerical lens  galaxy model.  Four  methods have
been  applied  to  determine  the  time  delay  between  the  two  QSO
components, giving  a mean estimate of  $\Delta t =  51\pm4$ days ($95
\%$ confidence  level).  This  is the fourth  optical time  delay ever
measured.  Adopting a $\Omega=0.3$, $\Lambda=0$ Universe and using the
mass model of \citet{Maller}, this time-delay estimate yields a Hubble
parameter    of   $H_{0}=52^{+14}_{-8}~{\rm    km}~{\rm   s^{-1}}~{\rm
Mpc^{-1}}$  ($95  \%$  confidence  level)  where  the  errors  include
time-delay as  well as model uncertainties.   There are time-dependent
offsets  between the  two  (appropriately shifted)  light curves  that
indicate the presence of external variations due to microlensing.

\end{abstract}


%

\section{Introduction}

Intensive observational studies  of  gravitationally lensed QSOs  have
been conducted   in the last few   years with the  aim  of determining
cosmological parameters, e.g.,  the  Hubble constant, $H_{0}$, and  to
study the dark-matter distribution  in lens galaxies.  In  particular,
there has been a  significant effort  to  measure time delays  between
lensed  QSO components to derive  $H_{0}$ with the method described by
\citet{Refsdal64}. Following a more than decade-long monitoring of the
double    QSO  0957+561  by   Schild  and  coworkers  \citep{Schild90,
Schild95}, \citet{Schild97}  and  \citet{Kundic} succeeded in  pinning
down  its time  delay  in  1997.   The same  year  saw  an even   more
impressive accomplishment of  \citet{Schechter} to  determine multiple
time   delays in  the  `triple  QSO'   PG1115+080  from a peak-to-peak
variation of  barely 0.15  mag.   Following this  demonstration of the
feasibility  of  measuring  time  delays  of  multiply imaged  QSOs an
ongoing   photometric monitoring was  initiated  at the Nordic Optical
Telescope  (NOT) in April 1998.   The program involves measuring  time
delays between  lensed QSO components in  as many systems as possible.
Our main target  during the first   year of monitoring was the  doubly
imaged radio source QSO B1600+434 \citep{Jackson} at redshift $z=1.59$
\citep{Fassnacht},   which is   gravitationally  lensed by  an edge-on
late-type galaxy   at    $z=0.41$   \citep{Jaunsen,Fassnacht}.     The
1.39\arcsec\ angular separation between the two  QSO images (labeled A
and  B in Fig.~\ref{decima}),  together  with the observed large  flux
variations  in the system \citep{Jaunsen},  makes  it well suited  for
time-delay  measurements.  Given  the   poor  knowledge of  the   mass
distribution in  dark-matter halos of   spiral galaxies, a  time-delay
measurement  in B1600+434  may not  provide  a  firm determination  of
$H_{0}$.  However, once $H_{0}$ has  been determined from other lensed
systems or   using other  methods,  the time  delay  of  B1600+434 can
provide new  constraints on  the distribution  of  mass in the various
components of spiral galaxies in general \citep{Maller}.

B1600+434 is very faint, with $I\sim22$ for the faintest QSO component
and $I=20.3$ for the  lensing galaxy.  Furthermore, the B-component is
substantially obscured  by the   lens (see  Fig.~\ref{decima}) and 
photometry of the image is non-trivial.  
The data have   therefore been
analyzed with advanced deconvolution techniques both in order to model
the light distribution  of the lensing galaxy  and to achieve accurate
photometry  of the QSO component blended  with  the lens nucleus.  The
time delay we present using the  deconvolution technique is the fourth
one measured at optical wavelengths, and  it  is the first of  our monitoring
program.  A  preliminary    report    was  presented      in
\citet{Hjorth99}.  The results presented here include more data points
and supercede  this  report.  Independently, a   time  delay has been
measured  at radio wavelengths with the  VLA during the same observing
season \citep{Koopmans}.


\section{Observations and Data Reduction}

Weekly observations  of  B1600+434 were carried out   at the  NOT from
April 1998 to November 1999,  except for a   short period (i.e.,  from
December 1998 to February 1999) where the object was below the horizon
at   the   NOT.  Three   different    instruments  were  used:  ALFOSC
(Andaluc\'\i  a  Faint Object Spectrograph),   HiRAC (High  Resolution
Adaptive    Camera) and the    stand-by camera  StanCam, equipped with
detectors   yielding  pixel scales   of   0\farcs188,   0\farcs107 and
0\farcs176  respectively.  The $I$   band was chosen  to  minimize the
effect of the  extinction of the B  component  by the lensing  galaxy.
The  total exposure times for each  data point were adjusted according
to the moon phase and typically ranged from 20 to 40 min, divided into
three dithered exposures.  The seeing  usually varied from 0\farcs7 to
1\farcs4,  with 0\farcs9  being the most    frequent value. A  typical
signal-to-noise ratio (S/N) of 100 was reached for the A component and
50  for the B  component.  Color terms were   determined for the three
detectors in order  to match the light curves  obtained with the three
instruments.

An automated pipeline employing routines in the IRAF/NOAO/CCDRED package has 
been developed in order to pre-process the CCD frames in an efficient and
homogeneous way. Fringe-correction  and cosmic-ray removal were
performed on the individual frames before combination.

\section{Photometry}
\label{sect:phot}

The photometric data consist  of  one stacked frame per epoch. 
All light curves are calculated relative   to  3 stars in the
field  that have calibrated $I$ band  magnitudes \citep{Jaunsen}.  Two
of these  stars, angularly close  to the  QSO, labeled  S1  and S2  by
\citet{Jaunsen}, are used to construct the Point Spread Function (PSF).

All the combined frames, one per epoch, are  deconvolved with the MCS
deconvolution algorithm   \citep{Magain}.  With this  technique,  many
frames  of  a single object  can  be simultaneously deconvolved.  This
procedure   combines  the total  S/N  of   all the frames  obtained at
different periods to determine  the light distribution of the extended
sources (galaxies) as   well as  the positions  of   the point sources
(QSOs), since these parameters do not vary with time.  The intensities
of the point sources,  however, are allowed to  vary from one image to
the other, hence producing    the  light curves.  This technique    is
particularly well suited for the  analysis of B1600+434 because of the
light contamination  of  the faint B  component by  the spiral lensing
galaxy.   Simultaneous  deconvolution of  all  the frames allows us to
derive  a     high signal-to-noise   numerical     galaxy model   (see
Fig.~\ref{decima}).  The derived galaxy  is in good  agreement with
the images of B1600+434 obtained with the Hubble Space Telescope (HST)
\citep{Maller}. 

Maps of the residuals for each  frame are used to check deconvolution
results.  These maps represent the  $\chi^2$  fit in each pixel; i.e.,
the match between the  deconvolved model image  (reconvolved with the PSF)
and the data.  By  inspecting  these residuals  we conclude that  the
photometry of both components was mainly limited by photon noise.

In order to check for errors introduced  by the PSF we deconvolve two
fainter stars in the field  that are not  used in the construction of
the PSF.  Light curves for two of these stars, labeled  S3 and S4, are
displayed in Fig.~\ref{lightcurve}.  Assuming that these stars are not
variable, we use  the standard deviation around  the mean  value as a
measure of the photometric accuracy, giving  1$\sigma$ errors of 0.026
mag  and 0.039  mag  for S3 and   S4 respectively.  For  each point we
subtract the photon noise  in quadrature and attribute the residual
error to the PSF.  For the QSO images  these estimated PSF errors are
added in quadrature to  the photon noise in order  to model  the total
errors in the photometry.

\section{Light curves}
\label{sect:lcurve}

The $I$-band  light curves   (Fig.~\ref{lightcurve}) contain  41  data
points  for each component.   As   predicted by lens theories,  these
light curves show  that A  is the leading  component.  A  feature in A
observed at JD~2451050 (September~1998) is repeated in B about 50~days
later.  In  June~1999 ($\sim$~JD~2451300) there  is  an increase in the
flux from both components. This increase is particularly strong in the
B component and  represents more than  one magnitude in less  than 100
days.   After the  intensity increase both  A and B  display several
sharp  short-term variations.  Unfortunately our  light curves are not
sufficiently well-sampled  during  this  period to  recover  the exact
variations. This suggests that there  are variations on time scales
faster than the sampling frequency.

The features   in the light  curves are  sufficiently  distinct that a
rough eye-ball estimate directly yields  an approximate time delay  of
about 50  days.  Although  41  brightness  measurements are  a fairly  small
amount of   data, application of  standard  techniques to measure time
delays yield fairly robust results.  We analyzed the light curves with
the four methods described below.

\section{Time delay measurements}

\subsection{The SOLA method}
\label{sect:pijpers}

The method of   subtractive, optimally-localized averages  (SOLA)  was
originally developed for solving  inverse problems  in helioseismology
\citep{PT94}. The   basic  idea of this  method  is  to construct  an
optimal solution,    taking into account   measurement errors,  of any
linear inverse problem by taking  linear combinations of the  measured
data.  After having  been  applied successfully  in helioseismology the
SOLA  method  has    found   application   in  image    reconstruction
\citep{Pij99}, in the reverberation mapping of active galactic nuclei
\citep{PW94}, and in the determination  of time delays between
lensed QSOs \citep{Pij97}.

In    the case of  lensed  quasar images, a transfer
function  which is  a Dirac delta  function  is positioned at the time
delay.  With this method both the delay and the relative magnification
of a pair of images are determined.  Applied to B1600+434 a time delay
of $55$ days and a flux ratio of $0.79$ is found.  Also determined is
a relative  offset, for instance due to  a contribution of the lensing
galaxy or  a foreground object  which has not  been  subtracted, and a
(linear)  drift  between   the images    which   could  occur due   to
microlensing events with long   time  scales.  Problems can occur   if
there  are  higher  order  relative drifts  such  as short  time-scale
microlensing events which cannot easily  be accounted for within  this
method which just uses determinations   of  low order moments of   the
transfer function.

The errors   are obtained  by  assuming  that  the photometric errors  are
uncorrelated  and follow a   Gaussian  distribution. These  propagated
errors give a conservative error estimate of  $\pm10$ days on the time
delay value, i.e., $\Delta t = 55\pm10$ days.

\subsection{The minimum dispersion method}
\label{sect:pelt}

For the combined  data set generated from  the A light curve and   
the shifted B light  curve, we    estimate the
dispersion  of  the scatter around  the  unknown mean curve.  The true
time  delay between the   images should be manifested  as  a minimum in  
the dispersion spectrum  (see \citet{Pelt96} for definitions and notation).
Both the  simplest string length type  statistic $D_2^2$ and the  smoothed  
dispersion   spectrum  $D_3^2$ (smoothing parameter $\delta = 10$ days) 
give a clear global minimum at $\Delta t \approx 48$ days.

The precision of this time delay value is estimated using a bootstrap
procedure.  We first construct the combined  light curve from A and
the time-delay shifted B  curve ($\Delta t  = 48$). This yields  a
reference curve by median smoothing. From the reference curve we build
bootstrap samples  by   adding  randomized errors.   For   each sample
(altogether $1000$) we  compute $D_3^2$ dispersion  spectra and find
corresponding dispersion minima.  The scatter of the dispersion minima
for    the    different    bootstrap   runs    is    significant  (see
Fig.~\ref{pelt2}), with a formal $1 \sigma$  error estimate of $ \pm 16$
days.

\subsection{Model fit method}
\label{sect:chifit}

We model the data  with  an arbitrary  continuous  light curve with  a
fixed and constant sampling.  For a given time delay value $\Delta t$,
we use $\chi^2$ minimization to compare the model curve to the A curve
and the appropriately time delay shifted and magnitude offset B curve.
Additionally, a  linear term in one  of the curves  can be included to
model long term microlensing effects such  as observed in QSO~0957+561
\citep{Schild91}.  The minimization procedure is  repeated for a range
of time delay values  (0 to 100) and the  $\chi^2$ minimum is obtained
with $\Delta t  = 49$ days and $\Delta  m = 0.66$ magnitudes (i.e.,  a
flux ratio  of 0.54).  An additional small  positive linear  term for
the A component further improves the $\chi^2$ value.

A bootstrap method is used to estimate the errors.   Two sets of 1000
curves are constructed with the same number of  data points as in the
data. One set is  constructed with the same time  sampling as that of
our measured data, and the other with a random time sampling.  Running
the  program on  these  simulated  curves  results  in a  $1  \sigma$
standard deviation   of 2 days on  the   set of  curves  with the same
sampling as our measured data, and of 7 days on the curves with random
sampling. We could  interpret this as an error  of 2  days internal to
the method since the sampling stayed  the same for each simulated data
set, and an error of  7 days on the  measured time delay independently
of the   sampling of  the data, but   this  needs  to be  investigated
further.

\subsection{Iterative modeling}
\label{sect:iter}
Assuming  that  the  additional  time dependent   offsets  between the
time-delay shifted curves are  caused by microlensing, a fourth method
based on iterative correction  of the flux  ratios is applied to  the
data.   We split the  light curves into several  parts (bins).  In one
case  we chose three  bins with  lengths determined  according  to the
apparent lengths of the offsets in a  certain directions (positive or
negative compared to the other curve).   In three   other cases  we
separate the curves into  2,3 and  4  bins with  an equal number of  data
points.  For a range of time-delay values (40 -- 65 days) we first fit
a  model  (using   the $\chi^2$  model  fit   method  described above)
independently to the separate  bins.   For each   bin and time   delay
value,  an additional  constant or   linear offset  is determined   to
improve the  fit.  Finally, new  time  delay values are  determined on
these modified  curves.  The time delays determined  in  this way turn
out to show very  little sensitivity to  the input values used, and to
the different splitting of the curves.  The results converge towards a
mean value  of $51$ days with  a standard deviation  ($1 \sigma$) of 2
days and magnitude   offsets  varying from   0.6  -- 0.87 mag  in  the
different bins.

\subsection{Results}

The four time-delay estimates obtained  from the different methods are
consistent with one  another  (see Table~\ref{timed}).  An  average of
the results gives a time delay estimate of $51$  days and a flux ratio
$A/B=1.50$. The flux  ratio at radio  wavelengths has been measured to
be $1.21$ \citep{Koopmans}.  If  we assume that the  radio flux is not
affected by reddening, the B-component is reddened by 0.22 magnitudes,
corresponding to  a factor of   1.23.  At a   redshift of 0.41,   this
corresponds to   $A_V\sim0.25$ mag, essentially   independent of   the
assumed  reddening law. 

Except from the iterative modeling, the  methods described above yield
substantial errors in the   time delay.  However, the  good  agreement
between  the  results  in  addition to   the  observed  time dependent
magnitude offsets in the  time-delay shifted curves (Fig.~\ref{shift})
suggest  that the large  errors  are mostly  due   to the presence  of
external variations, e.g.,   microlensing effects, rather than  to the
method used  to  determine the time delay.   SOLA  and the   model fit
method  correct for slow,    linear variations, but  not  higher order
effects.  Let   us consider a case  in  which the  brighter of the two
lensed  images undergoes many microlensing events  during  the time of
monitoring. Such short term variations would be indistinguishable from
measurement noise for these  methods.  They  may  fail to  converge or
produce a result that is biased at a level of the order of the errors.
With the  iterative method however,  we also correct  for higher order
effects.   The small errors from the  iterative method compared to the
other three    methods indicate the   presence  of such   higher order
external variations in the light curves.

If no assumptions are made on the short term offsets between the time
delay shifted curves we must use the conservative error estimate
of $\pm 10$ days ($1 \sigma$) on the time delay value. 
However, if we assume that these external variations are higher order
effects we can use the $1 \sigma$ error estimate of
$\pm 2$ days as found from the iterative method.

\section{$H_{0}$ estimate}

Assuming that  microlensing effects affect our $I$  band light
curves, a time delay of $51\pm4$ ($95 \%$ confidence level) days
has been estimated.   Applying the galaxy models from \citet{Maller},
and assuming a  $\Omega=0.3$  and $\Lambda=0$ Universe,   our measured
time     delay   is      consistent    with   a    Hubble    parameter
of $H_{0}=52^{+14}_{-8}~{\rm    km}~{\rm s^{-1}}~{\rm Mpc^{-1}}$ ($95 \%$
confidence level)  where both errors on  the model and  the time delay
are included.

In the  model of B1600+434, \citet{Maller} assume  a  constant M/L for
the disk and the bulge, and  a dark halo  that has the same center  and 
the same orientation as  the   disk.  In addition  they include   the mass
associated with the companion galaxy (see Fig~\ref{decima}) modeled as
an isothermal  sphere.  The  dark  matter  is  modeled as a  standard
Pseudo          Isothermal        Elliptical  Mass        Distribution
\citep[PIEMD]{Kassiola}, and the exponential  profiles of the disk and
the bulge are modeled by a two-point PIEMD,  also called the chameleon
profile \citep{Keeton, Hjorth}.   Two types  of solutions  are  found
with these models, one with  a large dark-matter halo core  radius,
and the  other with a  nearly  singular dark halo.  Applying  only the
solutions    with     a large   dark      halo   core   radius   give
$H_{0}=54^{+6}_{-5}~{\rm km}~{\rm  s^{-1}}~{\rm Mpc^{-1}}$.

For   comparison we also estimate   $H_{0}$  using a simple analytical
generalized isothermal   galaxy model  as  described by  \citet{Witt}.
This  method depends only on  the  time delay and  the observed images
positions.  Using our measured time delay and the image positions from
\citet{Maller}     measured  on    the     HST   image    we    obtain
$H_{0}=45^{+6}_{-5}~{\rm       km}~{\rm      s^{-1}}~{\rm   Mpc^{-1}}$
($\Omega=0.3$ and $\Lambda=0$) where errors come only from the time delay 
and the image positions.  This  is  only an  indicative estimate   of $H_{0}$
using a very simplified model.  Possible shear from ellipticity in the
lens galaxy  or mass contribution   from the  companion galaxy is  not
taken  into  account.  

Finally, as pointed out by  \citet{Romanowsky} and \citet{Witt}, 
the  time delay depends on  the density profile of  the lens galaxy.
Hence if the real density profile of the spiral galaxy is different
from our models the errors in $H_{0}$ will increase.

\section{Discussion}

The main goal  of  our optical monitoring  program at  the NOT  is  to
measure  time delays for  as many lensed QSOs as  possible in order to
better constrain the mass distribution in lens galaxies. A robust time
delay value  of $\Delta t =  51 \pm10$ (1  $\sigma$)  days and  a flux
ratio  of 0.69 have been measured  from $I$ band   light curves of our
first target B1600+434.  Assuming that our lightcurves are affected by
external variations the  time delay value  is constrained  to $51\pm4$
days  ($95 \%$ confidence  level).   This value  is consistent  with a
Hubble    parameter $H_{0}=52^{+14}_{-8}~{\rm km}~{\rm    s^{-1}}~{\rm
Mpc^{-1}}$  ($\Omega=0.3$,  $\Lambda=0$)     using  the   models    of
\citet{Maller}.  We   recall that possible    systematic errors due to
uncertainties in  the density profiles and  the degeneracy between the
relative contribution  of  disk, bulge and  halo  to the  mass in  the
spiral galaxy may increase the uncertainties in the estimated $H_{0}$.

As is evident  from Fig.~\ref{shift}, and as  measured by  the various
methods to determine  the time delay, there  are slow additional  time
dependent magnitude offsets between  the curves.  These variations are
$\sim$0.2 mag on time scales of a few months.  Since the QSO images of
B1600+434 pass through the lens  galaxy with  high optical depths  for
microlensing, the  offsets between the  two time-delay  shifted curves
(Fig.~\ref{shift}) are likely to be due  to microlensing of one of the
components. This interpretation is supported by the detection of 
microlensing  in light   curves  obtained   at  radio  wavelengths
\citep{Koopmans}, and implies  that  a significant fraction  of the
mass in the dark halo consists of massive compact objects. Simulations
of microlensing light curves must be carried out in order to determine
the lens  masses and source sizes   that could reproduce  the observed
variations   in our   optical   light curves   (see \citet{WP91}   and
\citet{SW98}).  Such a microlensing  analysis  will be published in  a
separate paper.

Microlensing  could also be partly responsible  for the sharp event at
$\sim$JD2451350  in  the light curve  of  the B  component.  Such high
magnification and   short  duration events  may    occur in  cases  of
microlensing by stars in random motion whereas the slow variations are
typical for microlensing events by the stars with velocities following
the bulk motion of the galaxy (e.g., Wambsganss \& Kundi\'c 1995).  We
note   however that much of  this   event  must be   due to  intrinsic
variations in the QSO   since  a  significant peak  is
detected for both components.

Although the temporal  sampling of our curves  does not allow to fully
disentangle high frequency  microlensing and  intrinsic variations, it
is sufficient  to follow lower  order variations, yielding robust time
delay estimates  fairly independently of  the statistical method used.
The time delay estimates measured with the  four different methods are
in  agreement with each other.   Furthermore, the  time delay value of
$47^{+12}_{-9}$   days  recently   estimated from   radio measurements
\citet{Koopmans}  is consistent with our   optical measurement.  There
are  thus  two independent data  sets,  one at  optical wavelengths ($I$
band)  and one at radio  wavelengths, hence affected in very different
ways by microlensing,  yielding the same time  delay value within  the
measurement errors.

The  remarkably strong variations in  this system makes it interesting
to study in more detail. The observed microlensing effects 
may  provide important  constraints   on the MACHO masses  in  the lens  
galaxy.    Moreover,   once  $H_{0}$ is    known
independently from other  lenses, or other  methods, an accurate  time
delay can be used to constrain the mass  distribution between the halo
and the bulge  of   the spiral  lens  galaxy.  In  particular,  a well
determined time  delay and flux ratio  can contribute to determine the
maximum allowed disk mass in the lens galaxy.

\acknowledgements

We thank  the  NOT Director Vilppu Piirola   for granting us observing
time for this project on a flexible  basis. We are especially grateful
to the dozens of visiting  observers at NOT   who have contributed  to
this  project by performing the   scheduled observations. This project
was conceived in 1997 while  JH, AOJ, and  JP were visiting scientists
at  the Center for Advanced Study  in Oslo. We  thank Rolf Stabell and
Sjur Refsdal for inviting us there and  for their kind hospitality. We
also  acknowledge stimulating conversations  with Frederic Courbin and
the  useful comments  from the referee.  The  project was supported in
part by  the Danish Natural  Science Research  Council  (SNF).  IB was
supported in  part   by contract  ARC94/99-178 ``Action de   Recherche
Concert\'ee de  la Communaut\'e Fran\c{c}aise  (Belgium)'' and  P\^ole
d'Attraction Interuniversitaire, P4/05 \protect{(SSTC, Belgium)}.

\clearpage


\figcaption[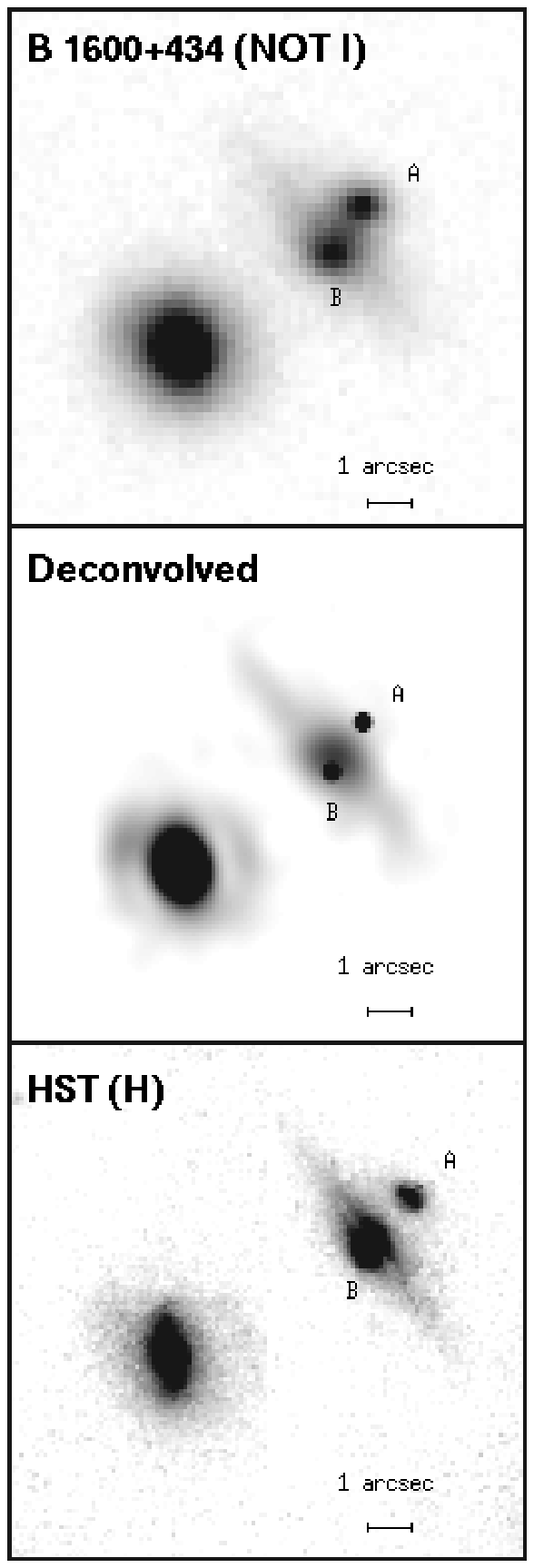]{{\it Top}: Stacked $I$-band images of $12\times12$ arcsec 
around B1600+434 with a total of $\sim 3.5$ hours of exposure and a seeing
FWHM = 1\farcs13.
{\it Middle}: The  image (FWHM = 0\farcs38)
as obtained from the simultaneous deconvolution of 33 frames. 
The main lensing spiral galaxy can be 
seen between the two QSO components A and B.  A neighbor galaxy 
is seen to the south east of the system.  North is up and East is
to the left. {\it Bottom}: The HST $H$ band image.
\label{decima}}

\figcaption[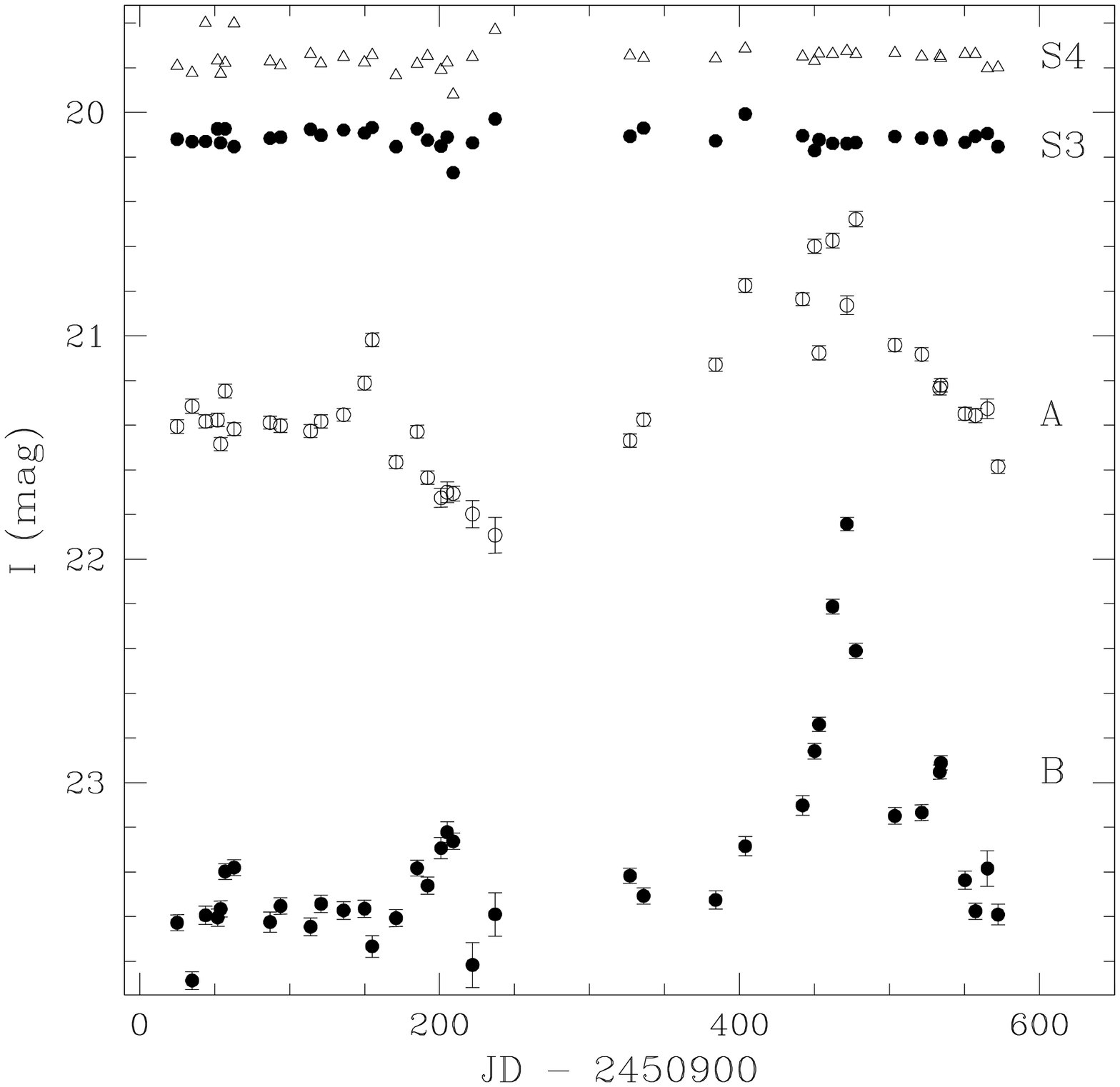]{
$I$-band light curves for B1600+434 (A and B component) and two 
reference stars in the field. The plotted value for B is $I$(mag)$+$1.5, 
and for S4 $I$(mag)$+$0.4. The magnitudes are calculated relative to
calibrated stars in the field. The error bars represent photon noise and
PSF errors measured as described in the text (\S~\ref{sect:phot}).
\label{lightcurve}}

\figcaption[f3.eps]{
Combined lightcurve from both components of B1600+434.
The curve from the B component is shifted forward in time by 51 days and scaled
with $-0.69$ mag. The magnitudes are calculated relative to
calibrated stars in the field. The error bars represent photon noise and
PSF errors measured as described in the text (\S~\ref{sect:phot}).
\label{shift}}

\figcaption[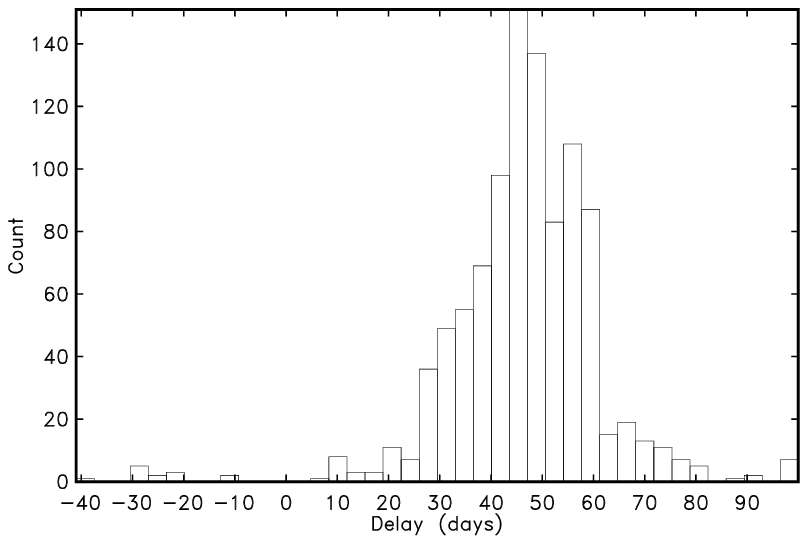]{Distribution of the time delays for $1000$ bootstrap 
runs with the minimum dispersion method (see \S~\ref{sect:pelt}).
\label{pelt2}}


\begin{deluxetable}{crr}
\tablecaption{Estimated time delays  and $I$-band magnitude differences 
for B1600+434 calculated with the four methods described in the 
text (\S~\ref{sect:lcurve}).\label{timed}}
\tablewidth{0pt}
\tablehead{
\colhead{} & \colhead{$\Delta$t (days)}   & \colhead{$\Delta m$ (mag)}
}
\startdata
SOLA       &  55$\pm$10  & 0.72$\pm$0.007 \\
Minimum dispersion &   48$\pm$16 & \nodata   \\
$\chi^2$ fit&  49$\pm$7 & 0.66$\pm$0.01\\
Iterative fit &  51$\pm$4  & 0.6--0.87 \\
\hline

\enddata
\end{deluxetable}


\begin{figure} 
\epsscale{0.4} 
\plotone{f1.eps} 
\end{figure}

\begin{figure} 
\epsscale{1.} 
\plotone{f2.eps} 
\end{figure} 

\begin{figure} 
\epsscale{1.} 
\plotone{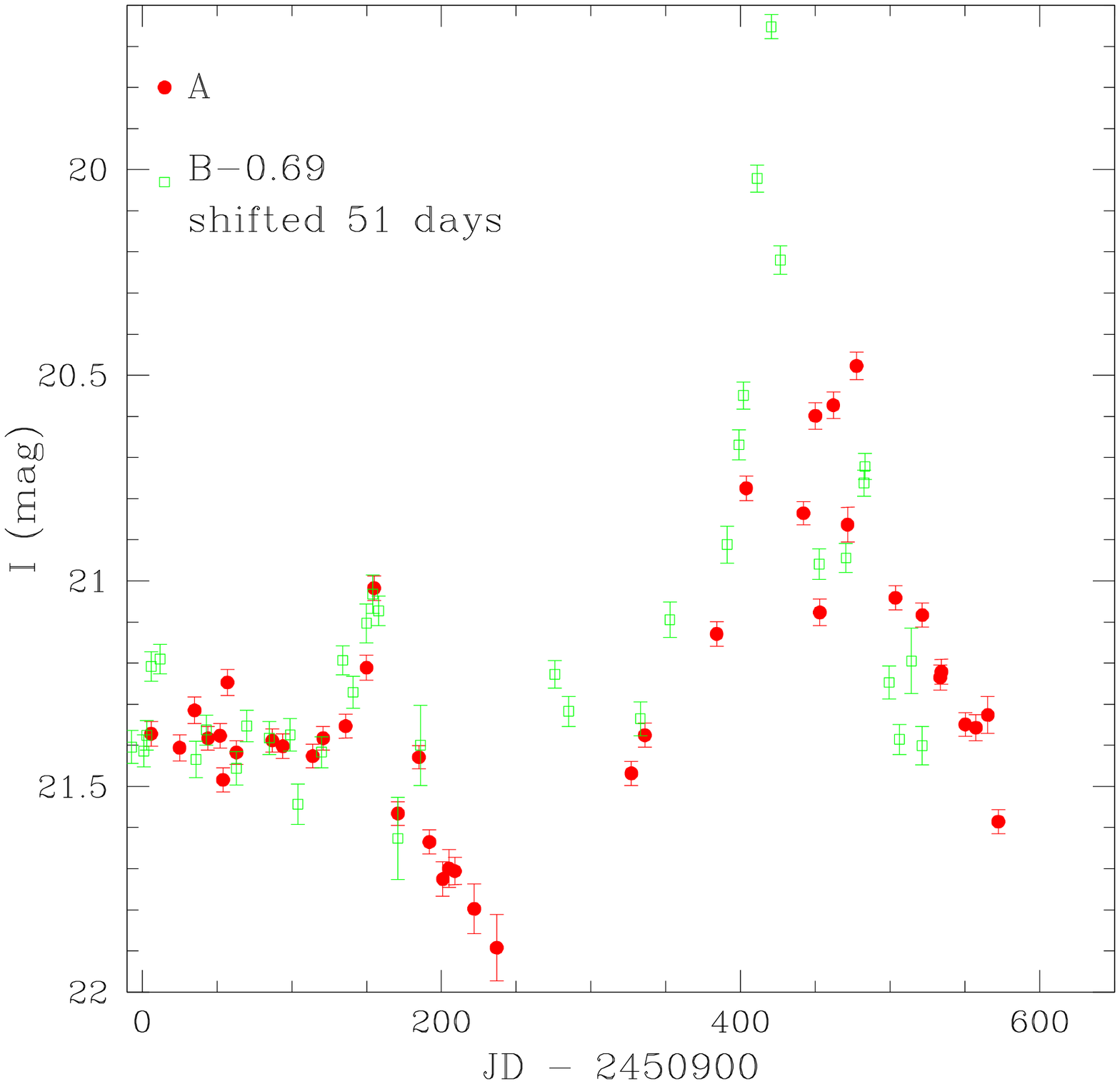} 
\end{figure} 

\begin{figure} 
\epsscale{1.} 
\plotone{f4.eps} 
\end{figure}


\begin{thebibliography}{}

\bibitem[Fassnacht \& Cohen(1998)]{Fassnacht} Fassnacht, C. D. \& Cohen, J. G. 1998, \aj, 115, 377
\bibitem[Hjorth et al.(1999)]{Hjorth99} Hjorth, J., Burud, I., Jaunsen, A. O.,
Andersen, M. I., Korhonen, H., Clasen, J. W., {\O}stensen, R. 1999, proceedings of the conference held in Boston 25-30 July, 1999:
           Gravitational Lensing: Recent progress and future goals. 
\bibitem[Hjorth \& Kneib(1999)]{Hjorth} Hjorth, J., \& Kneib, J.-P. 1999, \apj, submitted
\bibitem[Jackson et al.(1995)]{Jackson} Jackson, N.  et al., 1995, \mnras, 274, L25 
\bibitem[Jaunsen \& Hjorth(1997)]{Jaunsen} Jaunsen, A. O. \& Hjorth, J. 1997, \aap, 317, L39
\bibitem[Kassiola \& Kovner(1993)]{Kassiola} Kassiola, A., \& Kovner, I. 1993, \apj, 417, 450
\bibitem[Keeton \& Kochanek(1998)]{Keeton} Keeton, C. R., \& Kochanek, C. S. 1998, \apj, 495, 157
\bibitem[Koopmans et al.(2000)]{Koopmans} Koopmans, L. V. E, de Bruyn, A. G, Xanthopoulos, E., \& Fassnacht, C. D.  2000, \aap, 356, 391
\bibitem[Kundi{\'c} et  al.(1997)]{Kundic} Kundi{\'c}, T. et al.,  1997, \apj, 482, 75
\bibitem[Magain, Courbin \& Sohy(1998)]{Magain} Magain P., Courbin  F.,  \& Sohy  S.  1998,  \apj, 494, 472
\bibitem[Maller et al.(2000)]{Maller} Maller A. H., Simard, L., Guhathakurta, P., Hjorth, J., Jaunsen, A. O., Flores, R. A. \& Primack, J. R.  2000, \apj, 533, 194
\bibitem[Pelt et al.(1994)]{Pelt94}  Pelt J., Hoff, W., Kayser, R., Refsdal, S. \& Schramm, T.  1994, \aap, 286, 775
\bibitem[Pelt et al.(1996)]{Pelt96} Pelt J., Kayser R., Refsdal S. \& Schramm, T. 1996, \aap, 305, 97
\bibitem[Pijpers(1997)]{Pij97} Pijpers F. P. 1997, \mnras, 289, 933
\bibitem[Pijpers(1999)]{Pij99} Pijpers F. P. 1999, \mnras, 307, 659
\bibitem[Pijpers \& Thompson(1994)]{PT94} Pijpers F. P. \& Thompson M. J. 1994,\aap, 281, 231
\bibitem[Pijpers \& Wanders(1994)]{PW94} Pijpers F. P. \& Wanders I. 1994, \mnras, 271, 183 
\bibitem[Refsdal(1964)]{Refsdal64} Refsdal S. 1964, \mnras, 128, 295
\bibitem[Romanowsky \& Kochanek(1999)]{Romanowsky} Romanowsky, A. J. \& Kochanek, C. S. 1999, \apj, 516, 18
\bibitem[Schechter et al.(1997)]{Schechter} Schechter, P. L. et al., 1997, \apj, 475, 85 
\bibitem[Schild \& Thomson(1997)]{Schild97} Schild, R. E. \& Thomson, D. J.
1997, \aj, 113, 130
\bibitem[Schild \& Thomson(1995)]{Schild95} Schild, R. E. \& Thomson, D. J.
1995, \aj, 109, 1970
\bibitem[Schild \& Smith(1991)]{Schild91} Schild, R. E. \& Smith, R. C. 
1991, \aj, 101, 813
\bibitem[Schild(1990)]{Schild90} Schild, R. E 1990 \aj, 100, 1771
\bibitem[Schmidt \& Wambsganss(1998)]{SW98} Schmidt, R. \& Wambsganss, J. 1998 \aap, 335, 379
\bibitem[Wambsganss \& Paczy{\`n}ski(1991)]{WP91} Wambsganss, J. \& Paczy{\`n}ski, B. 1991, \aj, 102, 864 
\bibitem[Witt et al.(2000)]{Witt} Witt, H. J., Mao, S. \& Keeton, C. R. 2000, submitted, preprint astro-ph/0004069
\end{thebibliography}
\end{document}